\documentclass[twocolumn,           % Format:   preprint, twocolumn
               showpacs,            % Pacs  :   showpacs, noshowpacs
               nopreprintnumbers,   % Preprint: preprintnumbers, 	
               aps,                 % Society:
               prd,          	    % Style : pra,prb,prc,prd,pre,prl,
               a4paper,             % Size  :   a4paper, ...
               groupedaddress,      % Title :   groupedaddress, 
               unsortedaddress,
               nofootinbib,         % Foonote:  footinbib, nofootinbib 
               tightenlines,        % Remove additional spaces in a line
               floats               % Floating pictures and tables
               ]{revtex4}

\usepackage{graphicx}
\usepackage{amssymb,latexsym}
\usepackage[draft=false]{hyperref}

\begin{document}

%%--- DRAFTCOPY --------------------------------
%% Prints a large "DRAFT" diagonally across each page
%% Does not show up in TeXview
%% \typeout{Prints "DRAFT" on each page; does not show in TeXView}
% \special{!userdict begin /bop-hook{gsave 200 30 translate
% 65 rotate /Times-Roman findfont 216 scalefont setfont
% 0 0 moveto 0.90 setgray (DRAFT) show grestore}def end}
%%------------------------------------------------

%======================================%
%<<<<<<<<<<<< TITLE PAGE >>>>>>>>>>>>>>%
%======================================%

%\renewcommand{\topfraction}{0.99}
%\renewcommand{\bottomfraction}{0.99}

\title{Stochastic approaches to inflation model building}
\author{Erandy Ram\'{\i}rez and Andrew R.~Liddle}
\affiliation{Astronomy Centre, University of Sussex, 
             Brighton BN1 9QH, United Kingdom}
\date{\today} 
\pacs{98.80.Cq \hfill astro-ph/0502361}
\preprint{astro-ph/0502361}

%======================================%
%<<<<<<<<<<<<< ABSTRACT >>>>>>>>>>>>>>>%
%======================================%

\begin{abstract}
While inflation gives an appealing explanation of observed
cosmological data, there are a wide range of different inflation
models, providing differing predictions for the initial
perturbations. Typically models are motivated either by fundamental
physics considerations or by simplicity. An alternative is to generate
large numbers of models via a random generation process, such as the
flow equations approach. The flow equations approach is known to
predict a definite structure to the observational predictions. In this
paper, we first demonstrate a more efficient implementation of the
flow equations exploiting an analytic solution found by Liddle
(2003). We then consider alternative stochastic methods of generating
large numbers of inflation models, with the aim of testing whether the
structures generated by the flow equations are robust. We find that
while typically there remains some concentration of points in the
observable plane under the different methods, there is significant
variation in the predictions amongst the methods considered.
\end{abstract}

\maketitle
%======================================%
%<<<<<<<<<<<< MAIN TEXT  >>>>>>>>>>>>>>%
%======================================%

%%%%%%%%%%%%%%%%%%%%%%%%%%%%%%%%%%%%%%%%%%%%%%%%%%%%%%%%%%%%%%%%
\section{Introduction}

The impressive results from the Wilkinson Microwave Anisotropy Probe
\cite{wmap} have done much to improve the standing of inflation as the
leading paradigm for the origin of structure in the Universe.
However, they have not done much in the way of reining in the very
large number of viable inflationary models, as the uncertainty in the
key prediction of the spectral index $n$ remains significant (and
crucially encloses the special value of unity), and there is no sign
of primordial gravitational waves (specified by their ratio $r$
relative to density perturbations).

The large collection of inflationary models (see Ref.~\cite{LR} for
extensive reviews) has primarily been developed in an \textit{ad hoc}
manner, through selection of potentials motivated either by some
considerations from fundamental physics or by simplicity.  Typically
these potentials may have several parameters, meaning that the best
that observations can hope to do is constrain those parameters.  Only
if the potential is particularly tightly defined, for instance
$V(\phi) \propto\phi^6$, can it be ruled out by present observations.
Despite the \textit{ad hoc} way in which the collection of models has
been constructed, their collective predictions cover a fair part of
the $n$--$r$ observational plane, albeit not evenly.  It does not seem
appropriate, however, to interpret this as saying that inflation
models favour certain types of predictions.

An alternative approach is to throw away the idea of taking input from
fundamental physics and of enforcing simplicity (usually on the
potential $V(\phi)$ driving inflation), and instead seek to generate
models of inflation via some stochastic process, exploiting numerical
techniques where appropriate.  The archetypal such method is the
inflationary flow equations, introduced by Hoffman and Turner
\cite{HT} and generalized to high order by Kinney \cite{K} (see also
Refs.~\cite{HK,pwmap,EK}).  Intriguingly, models generated via the
flow equations exhibit a very clear structure in the observational
plane, primarily occupying the line $r \simeq 0$ or a diagonal locus
extending to positive $r$ and negative $n-1$.  The principal aim of
this paper is to consider whether or not such a structure is a robust
prediction of stochastically-generated inflation models, or whether it
is specific to the flow equations implementation.  As a by-product, we
provide a new implementation of the flow equations, and also explore
the origin of their observational prediction more closely.  We 
restrict ourselves to single-field inflation throughout.

\section{Flow equations revisited}

The flow equations take as their starting point a set of differential
equations linking a set of slow-roll parameters defined from the
Hubble parameter $H$. Following the notation of Kinney \cite{K}, these
are
\begin{eqnarray}
\label{e:eps}
\epsilon(\phi) & \equiv &  \frac{m_{{\rm Pl}}^2}{4\pi} \left( 
\frac{H'(\phi)}{H(\phi)} \right)^2 \,; \\
\label{e:srdef}
^{\ell}\lambda_{{\rm H}} & \equiv & \left( \frac{m_{{\rm Pl}}^2}{4\pi}
	\right)^\ell \, \frac{(H')^{\ell-1}}{H^\ell} \, 
	\frac{d^{(\ell +1)} H}{d\phi^{(\ell +1)}} \quad ; \quad \ell \ge 1 \,,
\end{eqnarray}
where primes are derivatives with respect to the scalar field. For
example, the parameter $^1\lambda_{{\rm H}}$ equals $(m_{{\rm
Pl}}^2/4\pi) H''/H$ and is often denoted $\eta(\phi)$. Using the
relation
\begin{equation}
\label{e:Nphi}
\frac{d}{dN} = \frac{m_{{\rm Pl}}^2}{4\pi} \, \frac{H'}{H} \, 
\frac{d}{d\phi} \,,
\end{equation}
where we define the number of $e$-foldings $N$ as decreasing with
increasing time, yields the flow equations
\begin{eqnarray}
\label{e:flow}
\frac{d\epsilon}{dN} & = & \epsilon(\sigma+2\epsilon) \,; \nonumber \\
\frac{d\sigma}{dN} & = & -5\epsilon\sigma - 12 \epsilon^2 +
	2(^2\lambda_{{\rm H}}) \,; \\
\frac{d(^\ell\lambda_{{\rm H}})}{dN} & = & \left[\frac{\ell-1}{2} \, \sigma
	+(\ell-2)\epsilon \right](^\ell\lambda_{{\rm H}})+^{\ell+1}\!\!\!
	\lambda_{{\rm H}} \; ; \; \ell \ge2 \,, \nonumber
\end{eqnarray}
where $\sigma \equiv 2(^1\lambda_{{\rm H}})-4\epsilon$ is a convenient 
definition. 
 
As pointed out in Ref.~\cite{flow}, these equations actually have
limited dynamical input from inflation, since in the form $d/d\phi$
they are a set of identities true for any function $H(\phi)$, and the
reparametrization to $d/dN$ modifies only the measure along the
trajectories, not the trajectories themselves. In that light it seems
surprising that they can say much about inflation at all, but it turns
out that the flow equations can be viewed as a (rather complicated)
algorithm for generating functions $\epsilon(\phi)$ which have a
suitable form to be interpreted as inflationary models \cite{flow}. In
the following section we will compare their results with more direct
ways of generating such functions, but in the meantime we will explore
the flow equations themselves further.

\subsection{A new numerical implementation of the flow equations}

The standard implementation of the flow equations \cite{K} decides a
truncation level for the hierarchy, sets ranges for the slow-roll
parameters within which they are randomly selected, and integrates the
flow equations either until the end of inflation, $\epsilon = 1$, or
until a stable late-time attractor is reached.  In the former case the
equations are then integrated backwards for a suitable number of
$e$-foldings (either a fixed number such as 50, or one also randomly
chosen within a range \cite{K}) where the observational quantities
$n$, $r$, and possibly others are evaluated and plotted. If a
late-time attractor is reached the observables are read off at that
point.

Here we use a new and more efficient implementation of the flow
equations, exploiting the fact that the flow equations have an
analytic solution discovered in Ref.~\cite{flow}. This is simply a
polynomial in $H(\phi)$:
\begin{equation}
H(\phi)=H_0 \left[ 1+A_1 \, \frac{\phi}{m_{{\rm 
Pl}}}+...+A_{M+1}\left(\frac{\phi}{m_{{\rm Pl}}}\right)^{M+1} \right]
\,.
\end{equation}
The coefficients $A_i$ can be written in terms of the initial values
of the slow-roll parameters as
\begin{eqnarray}
A_1=-\sqrt{4\pi\epsilon_0} \,;\hspace*{0.9cm} & \quad &
A_2=\pi(\sigma_0+4\epsilon_0)\,;\nonumber\\
A_3=-\frac{4\pi^{3/2}}{3} \; \frac{^2\lambda_{{\rm H},0}}
{\epsilon_0^{1/2}}\,; &\quad &
A_4=\frac{2\pi^2}{3} \; \frac{^3\lambda_{{\rm H},0}}
{\epsilon_0}\,;\\
A_5=-\frac{4\pi^{5/2}}{15} \; \frac{^4\lambda_{{\rm H},0}}
{\epsilon_0^{3/2}}\,; & \quad &
A_6=\frac{4\pi^3}{45} \; \frac{^5\lambda_{{\rm H},0}}
{\epsilon_0^2}\,;\nonumber\\
A_7=\frac{8\pi^{7/2}}{315} \; \frac{^6\lambda_{{\rm H},0}}
{\epsilon_0^{5/2}}\,. \nonumber 
\end{eqnarray}
We use the ranges specified in Ref.~\cite{K} to randomly choose those
values:
\begin{eqnarray}
\epsilon_0 & =& [0,0.8]\,;\nonumber\\
\sigma_0 &=& [-0.5,0.5]\,;\nonumber\\
^2\lambda_{{\rm H},0}&  =& [-0.05,0.05]\,;\\ 
^3\lambda_{{\rm H},0}& = &[-0.005,0.005]\,;\nonumber\\
&...&\nonumber\\
^7\lambda_{{\rm H},0} &=& 0\,, \nonumber
\end{eqnarray} 
where the last closes the hierarchy. To allow direct comparison with
Kinney's work, we use these ranges throughout, though one expects the
results to be at least modestly dependent on the assumptions made here
\cite{K}. However we take the equations to sixth-order, one order higher 
than Kinney's main results, as one of the methods we will compare with 
later can only be implemented for even orders. As already shown by
Kinney, and separately verified by us, such a change in order has 
negligible impact on the flow equation predictions. We have carried 
out flow analyses at fifth and eighth orders as well as those displayed here.

Although this solution is analytic, there is still the need for one
integration in order to determine the number of $e$-foldings as a
function of $\phi$, from Eq.~(\ref{e:Nphi}).  However this is just a
single equation to be integrated, regardless of the order to which the
flow equations are taken.

\begin{figure*}
\centering
\includegraphics[width=7.5cm]{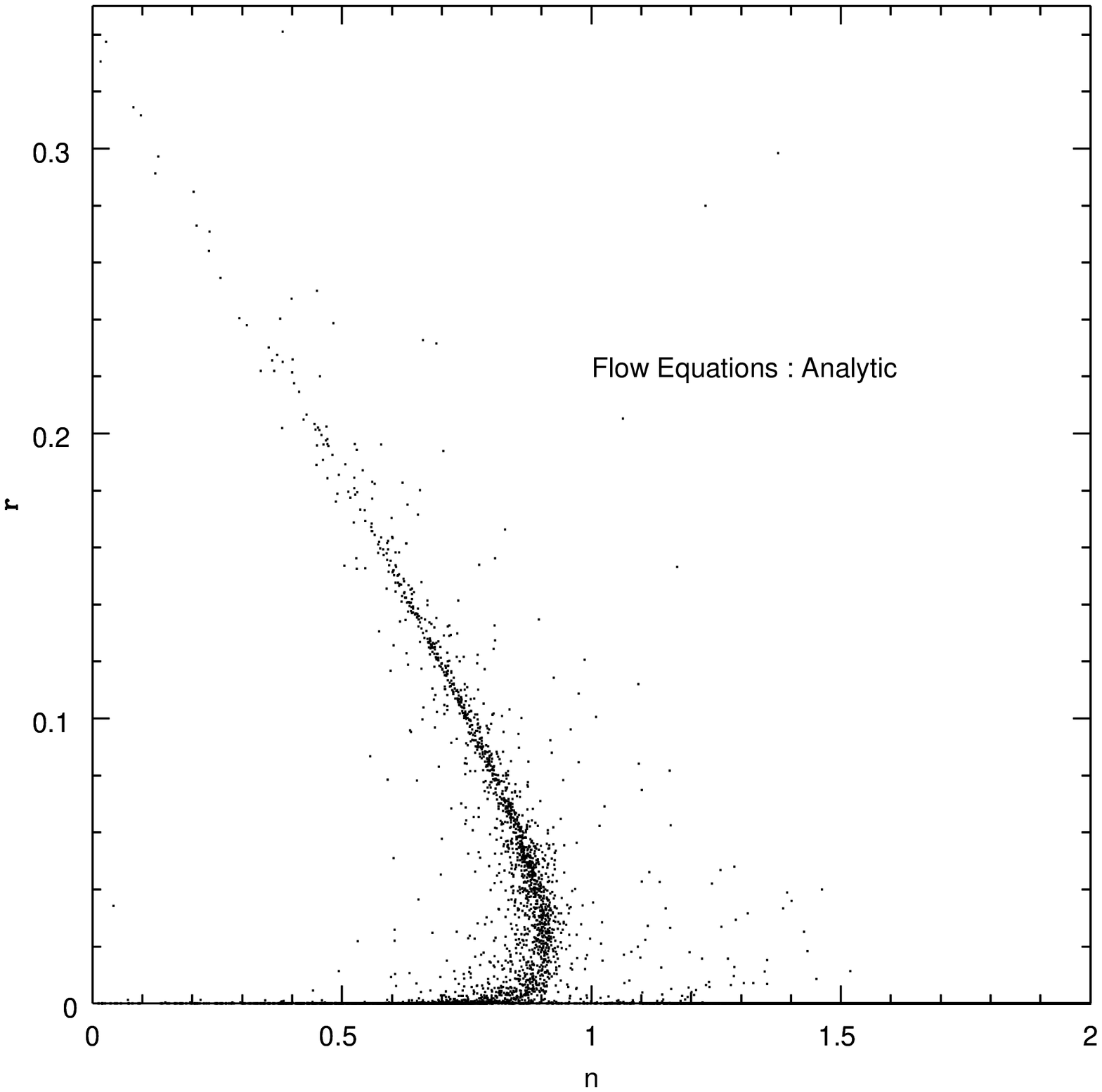} \hspace*{1.6cm}
\includegraphics[width=7.5cm]{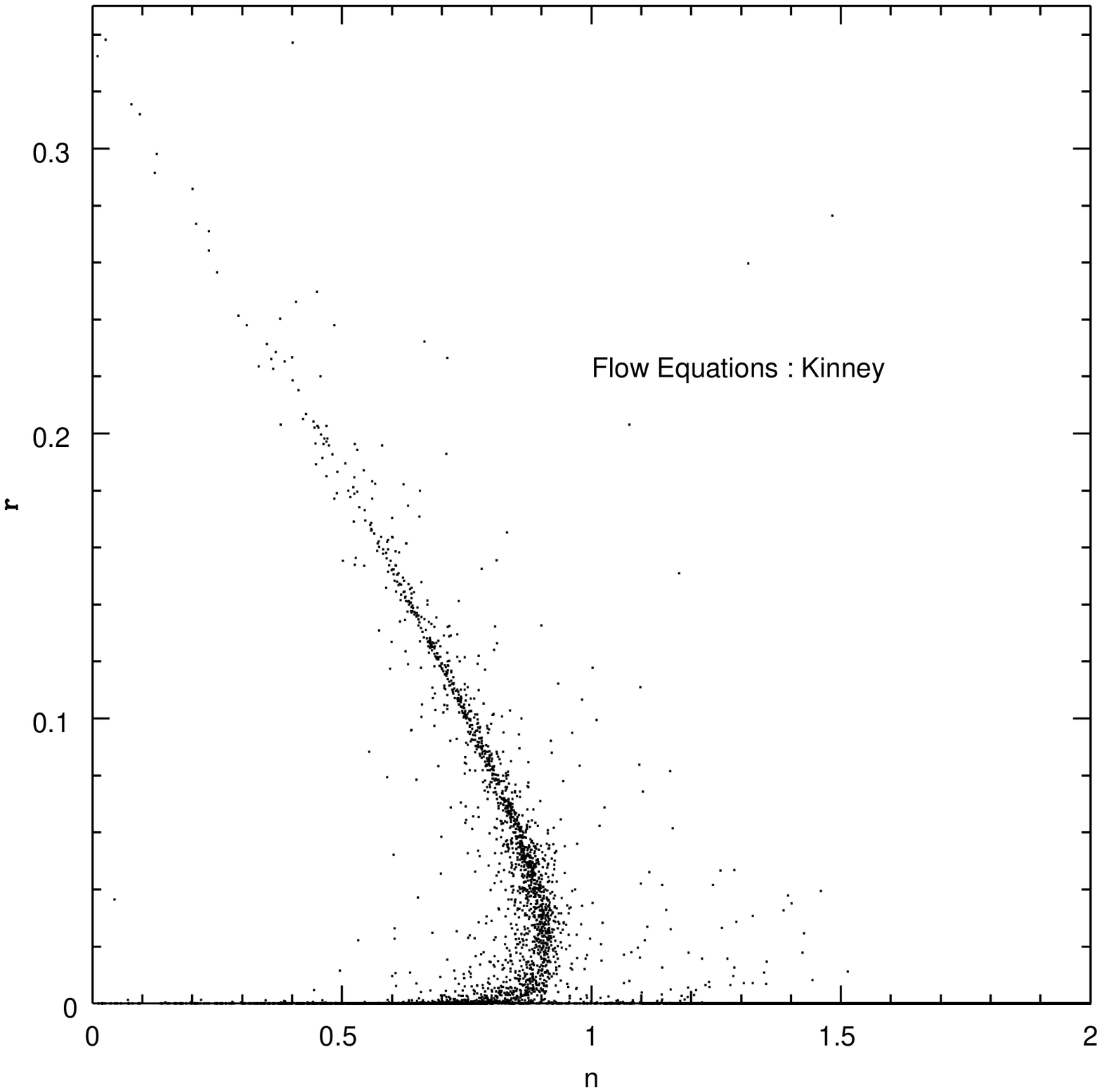}
\caption{Observables at second order for 40000 initial conditions. The
two figures are essentially identical, and reproduce the results of
Fig.~1 in Ref.~\cite{K} though at lower point density.}
\label{fig1}
\end{figure*}

Figure~\ref{fig1} shows the results from the flow equations, with the
left-hand panel showing our new implementation based on the analytic
solution, and the right-hand panel the traditional multi-equation
implementation. The same sequence of 40,000 initial conditions was
used in each case, with most of the points finishing at very small
$r$. The values of the spectral index $n$ and tensor-to-scalar ratio
$r$ were obtained using second-order expressions and the conventions
of Ref.~\cite{K}. As expected, the diagrams are essentially identical
point-by-point, though some minor differences occur from the way our
implementations differ at very small values of $\epsilon$.

The diagram shows a clear and by now well-known structure first
noted by Hoffman and Turner \cite{HT}; in addition
to the majority of the points at small $r$, there is a swathe of
models running in a tightly-defined strip given approximately by $r =
0.3(1-n)$, which is close to but not exactly the same as the power-law
inflation condition \cite{K}. While Kinney has been careful not to
overinterpret the tendency of points to lie in this vicinity, noting
that the measure on initial conditions is unknown, the results are
often used to indicate where typical inflationary models might lie
(e.g.~Ref.~\cite{pwmap}).  Our main aim in this paper is to
investigate the robustness of this structure.

\subsection{Integration direction}

First however we investigate in a little more detail how the structure
arises.  Kinney's main classification of trajectories is into those
reaching the late-time attractor (which all have $r \rightarrow 0$ and
$n>1$, corresponding to the field asymptoting into a non-zero minimum
of the potential), those trajectories where inflation ends with
$\epsilon = 1$, and the rejected set of trajectories which are unable
to sustain sufficient inflation. We make a further division of those
trajectories where inflation ends, into those where more than 50
$e$-foldings of inflation were obtained from the initial point (meaning
that the location where the observables were read off was reached by
forwards integration from the initial point), and those where less
than 50 $e$-foldings were obtained from the initial point, so that the
point corresponding to the observables is effectively obtained by
integrating backwards in time from the initial condition.

\begin{figure}
\centering
\includegraphics[width=7.5cm]{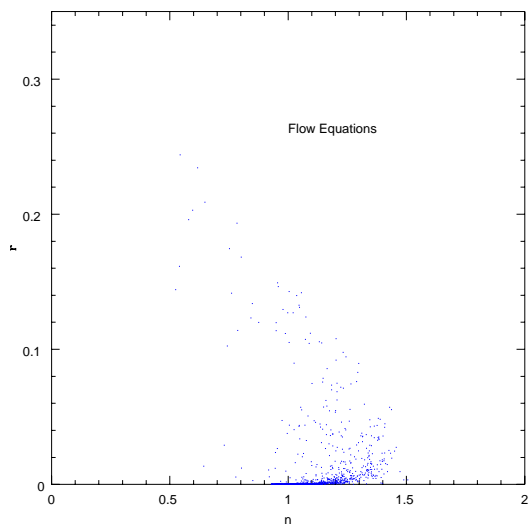}
\caption{Points in parameter space finishing inflation with $50$
$e$-foldings or more in the forwards integration.}
\label{fig2}
\end{figure}
 
We find that the vast majority of random initial conditions which end
inflation do so before $50$ $e$-foldings are achieved, in proportion
roughly 35 to 1. In Figure~\ref{fig2} we show only points obtained
by forwards integration; as there are so few of these we increased the
number of initial conditions tested up to 600,000, so as to have a
greater density of points (1225 in all, corresponding to just $0.2
\%$ of all initial conditions) than in Fig.~\ref{fig1}. We find that
those points correspond mainly to the region of parameter space where
$n>1$, and do not correspond to the main swathe of points seen in
Fig.~\ref{fig1}. That swathe is therefore generated entirely from
initial conditions that have to be integrated backwards in time. The
change of integration direction is significant, because the critical
point structure of the system is different in the inverse time
direction (the flow will typically be to the peaks of the potential
rather than the troughs). Although the backwards integration will pick
out those critical points, they need not represent a distribution that
might have arisen from whatever mechanism generated the initial
conditions.

However even the backwards integration does not explain the flow
equations structure, because the only stable critical points under
backwards integration are at $r \rightarrow 0$ and $n<1$. The diagonal
swathe is not stable, and, as already shown in Refs.~\cite{HT,K}, the
trajectories evolve along it, typically following quite complicated
paths when projected on the $n$--$r$ plane. Given that the
location of the swathe is known, it would be nice to explain that
behaviour via an analytic argument, but we have not been able to find
an analytic approximation that makes clear why points are able to
spend a long time in those parts of the plot.

\section{New approaches to stochastic generation  of inflation models}

\label{s:3}

We now turn to alternative methods of generating random inflation
models to contrast with the flow equations. As we have seen, the flow
equations are equivalent to a Taylor expansion of the function
$H(\phi)$, which uniquely specifies the inflationary
trajectory. However it is not the only way to do so.  Single-field
inflation models can also be uniquely specified either by giving the
potential $V(\phi)$ or, as noted in Ref.~\cite{L94}, by specifying the
function $\epsilon(\phi)$.

In order to investigate the robustness of the flow equation
predictions, one should therefore compare its results with those from
expansions of these alternative functions, as there is no reason to
prefer one over another. We will consider Taylor expansions of both
$\epsilon(\phi)$ and $V(\phi)$, and additionally consider a Pad\'e
approximant expansion of $\epsilon(\phi)$. In each case we take
the randomly-chosen coefficients to correspond to the same ranges of
slow-roll parameters used by Kinney for his flow equations analysis~\cite{K}.

If one of the three functions $H(\phi)$, $V(\phi)$ and $\epsilon(\phi)$ is 
specified, then the equivalent form of the others can readily be obtained. 
However once the functions are truncated as expansions at a given order, this 
correspondence breaks down, e.g. a sixth-order truncation of $H(\phi)$
does not correspond to a sixth-order truncation of $\epsilon(\phi)$. In 
carrying out these expansions, we are therefore investigating
different subsets of the general slow-roll inflation model. Even if the 
expansions were taken to infinite order the correspondence between models will 
only be obtained provided each series is within its radius of convergence, which 
is not guaranteed. On the
other hand, by choosing the same initial values for the slow-roll 
parameters in each case, we are ensuring that the functional forms at
the initial point are sampled from the same distribution. 

\subsection{$\epsilon(\phi)$ as the fundamental input}

The general strategy is similar to the new analytic approach to the
flow equations, in that we choose coefficients randomly to generate a
function $\epsilon(\phi)$, and then numerically integrate to determine
the number of $e$-foldings supported. However at this point we do have
to mention one drawback of using $\epsilon(\phi)$, which is that when
the function crosses zero (i.e.~into the unphysical region) it
typically does so linearly which means it does so in a finite number
of $e$-foldings. By contrast, the $\epsilon(\phi)$ generated from
either the Hubble parameter or the potential always approaches zero
quadratically, generating an infinite number of $e$-foldings. The
$\epsilon(\phi)$ expansions are therefore unable to generate points
corresponding to the late-time attractor.

\subsubsection{Taylor series expansion}

The simplest expansion we can make is a Taylor series
\begin{equation}
\epsilon(\phi) = \sum_{i=0}^K a_i (\phi/m_{{\rm Pl}})^i \,,
\end{equation}
where we assume $\phi=0$ initially. The higher slow-roll parameters
can all be obtained by differentiating this function, and so we can
determine the coefficients by randomly selecting the slow-roll
parameters within the same ranges as for the flow equations using
those relations. The relations, taking the expansion to sixth-order, are 
\begin{eqnarray}
a_0 &=& \epsilon_0 \,; \nonumber\\ 
a_1& =&
-\sqrt{4\pi\epsilon_0}(\sigma_0+2\epsilon_0)\,,\nonumber\\ a_2 &=&\pi
\left[\sigma_0^2-2\epsilon_0\sigma_0-12\epsilon_0^2+ 4(^2\lambda_{{\rm
H},0}) \right]\,;\nonumber\\ 
a_3& =&
\frac{4\pi^{3/2}}{3\sqrt{\epsilon_0}} \left[ -3(^2\lambda_{\rm
H_0})\sigma_0+2\epsilon_0(^2\lambda_{{\rm H},0})+6\epsilon_0\sigma_0^2
\right.  \\ && \left. + 21\sigma_0\epsilon_0^2
+12\epsilon_0^3-2(^3\lambda_{{\rm H},0}) \right] \,;\nonumber\\ 
a_4&=& -\frac{\pi^2}{3\epsilon_0} \left[ -12(^2\lambda_{\rm
H_0})^2+88\epsilon_0\sigma_0(^2\lambda_{{\rm H},0})+160\epsilon_0^2
(^2\lambda_{{\rm H},0}) \right. \nonumber \\ && 
-8\sigma_0(^3\lambda_{{\rm H},0}) +4\epsilon_0(^3\lambda_{{\rm
H},0})-45\epsilon_0^2\sigma_0^2-3 12\epsilon_0^3\sigma_0 \nonumber \\
&& \left. -384\epsilon_0^4+6\epsilon_0\sigma_0^3- 4(^4\lambda_{{\rm
H},0})\right] \,;\nonumber\\ 
a_5& =& 
-\frac{2\pi^{5/2}}{15\epsilon_0^{3/2}} \left[-240\epsilon_0^4
\sigma_0+40(^2\lambda_{{\rm H},0})(^3\lambda_{{\rm H},0})+
480\epsilon_0^3\sigma_0^3
\right. \nonumber \\ && \left.
+380(^2\lambda_{{\rm
H},0})\epsilon_9^2\sigma_0-140(^3\lambda_{{\rm
H},0})\epsilon_0\sigma_0-260(^3\lambda_{{\rm H},0})\epsilon_0^2
\right. \nonumber \\ && \left.
-80(^2\lambda_{{\rm H},0})\epsilon_0\sigma_0^2+1360(^2\lambda_{{\rm
H},0})\epsilon_0^3-200(^2\lambda_{{\rm H},0})^2\epsilon_0
\right. \nonumber \\ && 
-1440\epsilon_0^5+135\epsilon_0^2\sigma_0^3
+4(^5\lambda_{{\rm H},0})
+10(^4\lambda_{{\rm H},0})\sigma_0
\nonumber \\ && \left.
-4(^4\lambda_{{\rm H},0})\epsilon_0 \right] \,;\nonumber\\
a_6& =& 
\frac{\pi^3}{45\epsilon_0^2} \left[-840(^2\lambda_{{\rm 
H},0})^2\epsilon_0\sigma_0+
120(^2\lambda_{{\rm H},0})(^4\lambda_{{\rm H},0})
\right. \nonumber \\ && \left.
+5040\epsilon_0^3(^3\lambda_{{\rm H},0})
-768(^4\lambda_{{\rm H},0})\epsilon_0^2-
8(^5\lambda_{{\rm H},0})\epsilon_0
\right. \nonumber \\ && \left.
+1920(^2\lambda_{{\rm H},0})^2\epsilon_0^2
-1920\epsilon_0^4(^2\lambda_{{\rm H},0})
-408(^4\lambda_{{\rm H},0})\epsilon_0\sigma_0
\right. \nonumber \\ && \left.
+10080(^2\lambda_{{\rm H},0})\epsilon_0^3\sigma_0
+1380(^3\lambda_{{\rm H},0})\epsilon_0^2\sigma_0
\right. \nonumber \\ && \left.
+4140(^2\lambda_{{\rm H},0})\epsilon_0^2\sigma_0^2
-300(^3\lambda_{{\rm H},0})\epsilon_0\sigma_0^2
+8(^6\lambda_{{\rm H},0})
\right. \nonumber \\ &&  \left.
+80(^3\lambda_{{\rm H},0})^2+135\epsilon_0^2\sigma_0^4
-2340\epsilon_0^3\sigma_0^3-17640\epsilon_0^4\sigma_0^2
\right. \nonumber \\ && 
-1520(^2\lambda_{{\rm H},0})(^3\lambda_{{\rm H},0})\epsilon_0
+24(^5\lambda_{{\rm H},0})\sigma_0
\nonumber \\ && \left.
-33120\epsilon_0^5\sigma_0-
14400\epsilon_0^6 \right] \,.\nonumber
\end{eqnarray}

We are then able to solve the model using a single integration to find the 
$N(\phi)$ relation, as in our new flow equations approach. We then apply the 
same tests to the models thus generated: do they
lead to an adequate number of $e$-foldings, do they require backwards
integration to achieve 50 $e$-foldings, and if satisfactory where do
they lie in the observational plane?

\subsubsection{Pad\'e approximant expansion}

A Pad\'e approximant is an alternative to a Taylor expansion, which
typically exhibits better convergence properties. It is formed of a
ratio of two polynomials, which may have the same or different orders:
\begin{equation}
\epsilon(\phi) = \frac{\sum_{i=0}^M a_i (\phi/m_{{\rm Pl}})^i}{1+\sum_{i=1}^N 
b_i (\phi/m_{{\rm Pl}})^i} \,.
\end{equation}
There is a one-to-one correspondence between Pad\'e approximants and
Taylor expansions of the appropriate order ($K = M+N$); those
expansions then agree near the origin of the expansion but differ as
the expansion parameter, in this case $\phi/m_{{\rm Pl}}$, becomes of
order one, as is typical in single-field inflation models.

We generate the Pad\'e approximants from the Taylor series using a
routine from Numerical Recipes \cite{NumRec}; this routine is
restricted to $N=M$ which is why we chose a sixth-order Taylor
expansion above. Having generated $\epsilon(\phi)$ in this way, we
proceed as before. One additional caveat is that Pad\'e approximants
asymptote to constant values; this corresponds to power-law inflation
but is somewhat artificial and so we exclude points which require
backwards integration to generate 50 $e$-foldings, and which tend to a
constant asymptote between zero and one.

\begin{figure*}
\centering
\includegraphics[width=7.5cm]{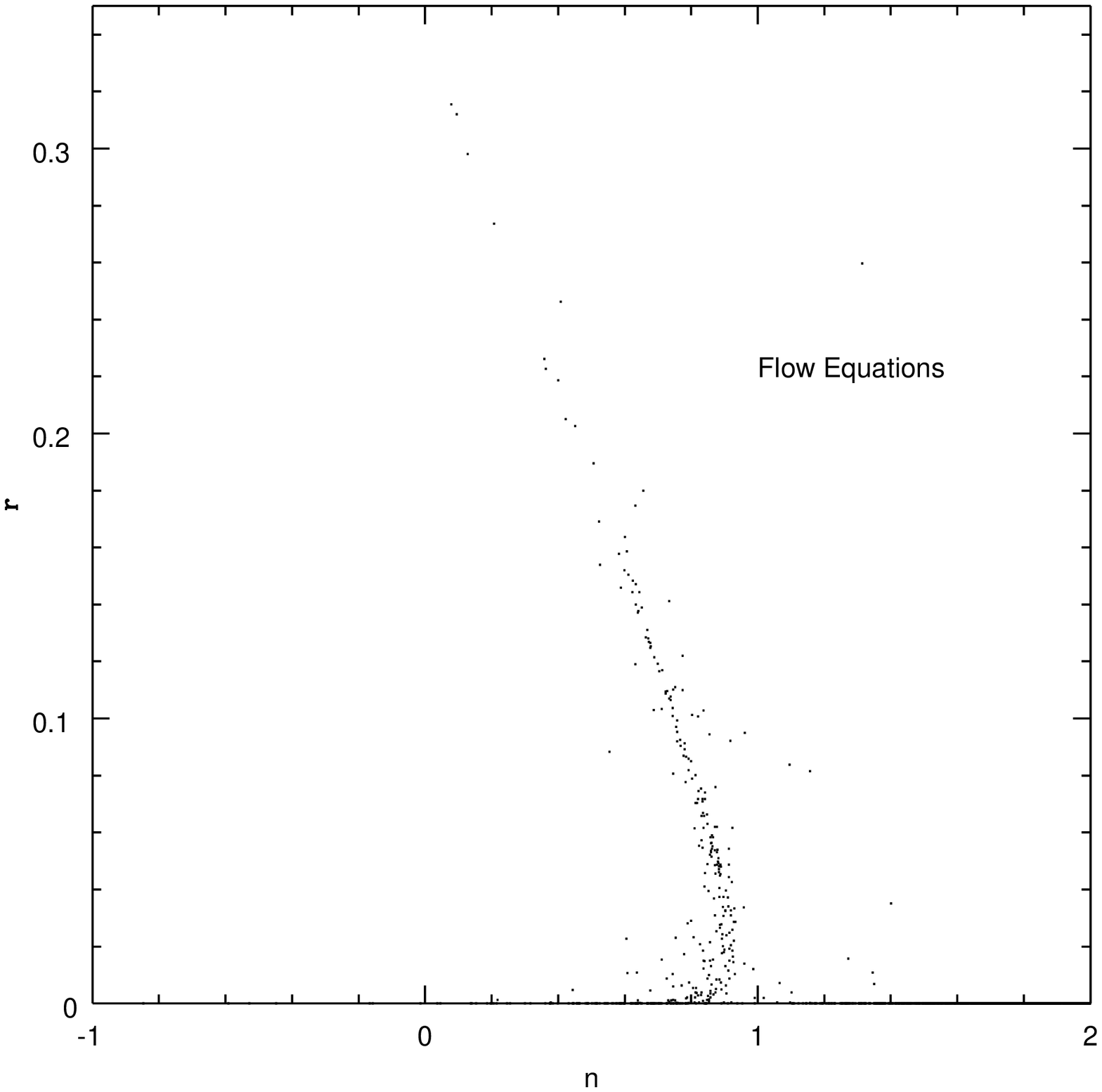}
\includegraphics[width=7.5cm]{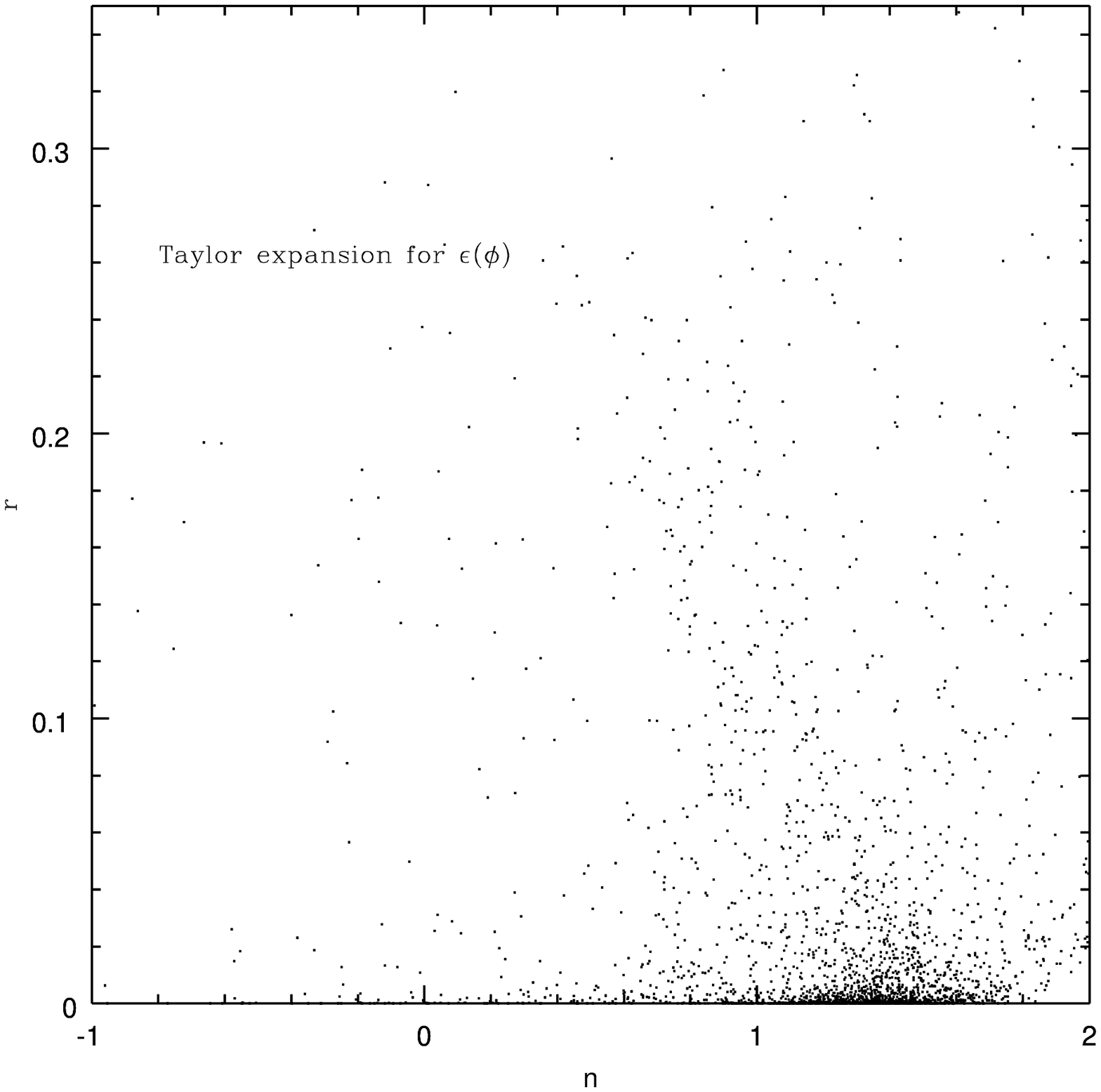}\\
\includegraphics[width=7.5cm]{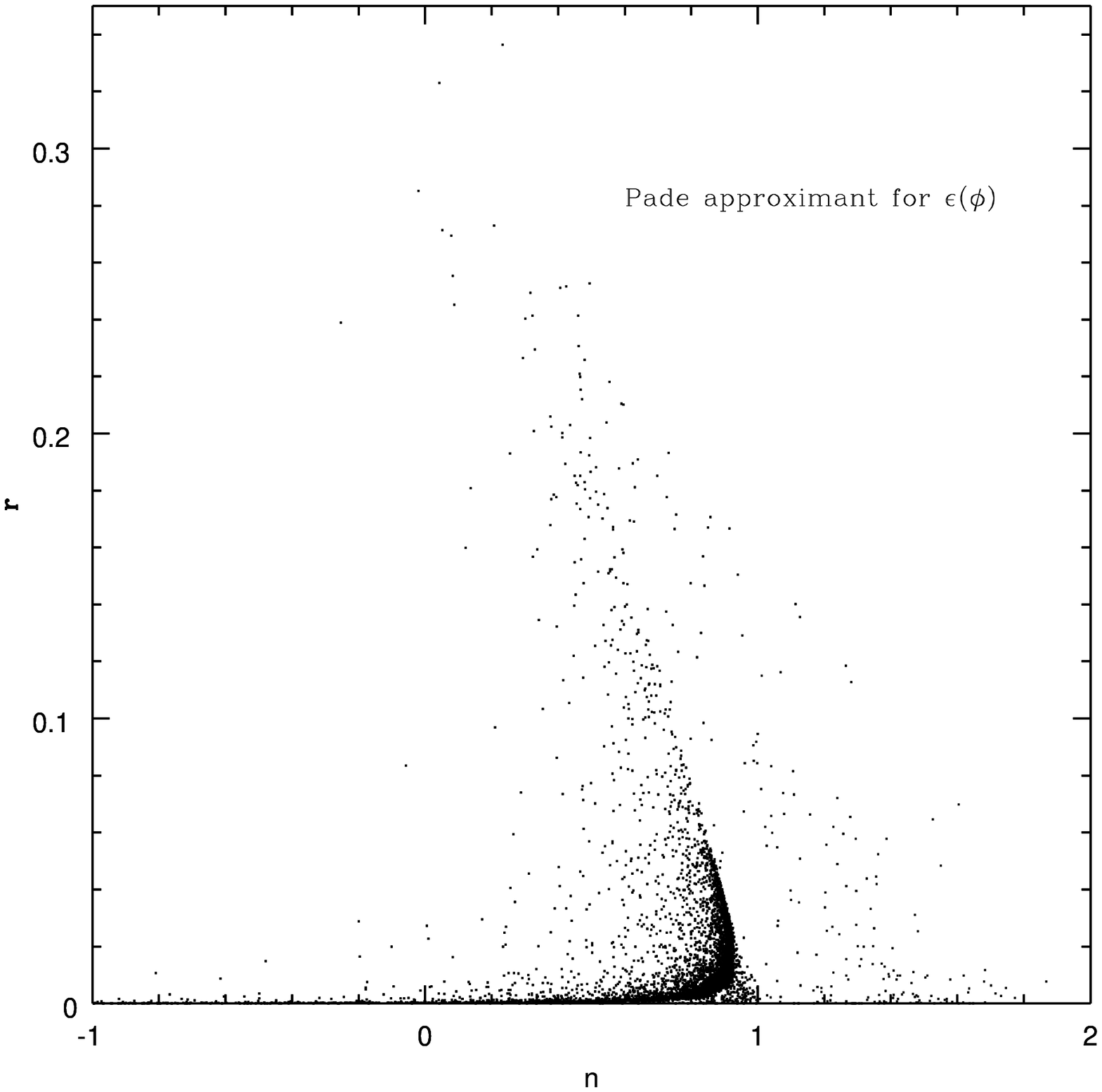}
\includegraphics[width=7.5cm]{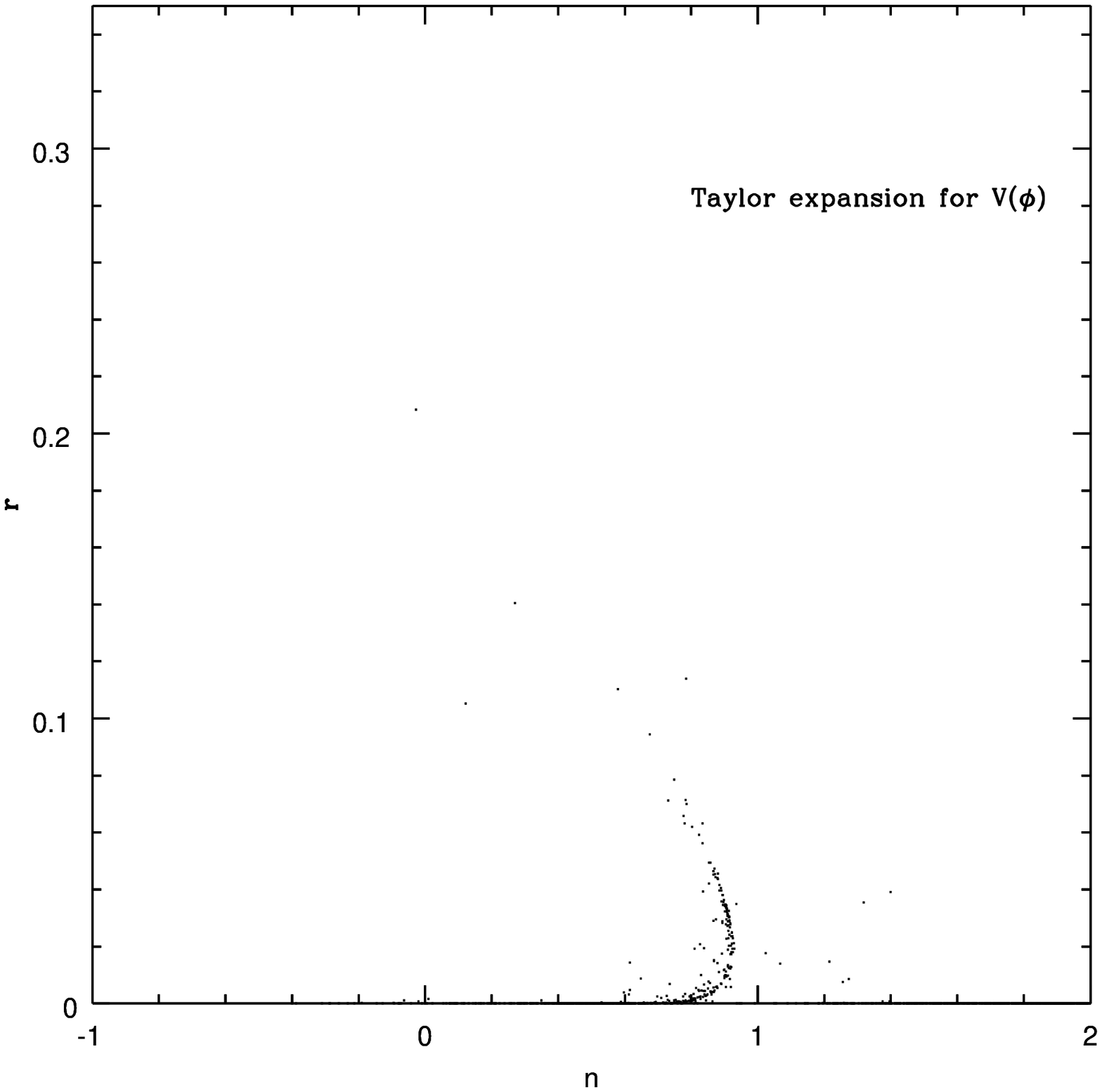}
\caption{The distribution in the observable plane for each of the four
methods discussed in this paper, in each case for 6000 accepted
initial conditions. On the top left are results for the flow equations, to
be compared with the Taylor expansion for $\epsilon(\phi)$ (top
right), the Pad\'e approximant for $\epsilon(\phi)$ (bottom left) and
the Taylor expansion for the potential (bottom right).}
\label{fig3}
\end{figure*}

\subsection{$V(\phi)$ as the fundamental input}

We can also generate the observables by Taylor expanding the potential
and integrating the $e$-foldings relation to find the values of $r$ and
$n$ $50$ $e$-foldings before the end of inflation. A similar expansion
was used to investigate inflationary dynamics in Ref.~\cite{MLL},
but that paper delineated parameter space regions rather than
generating ensembles of models. 

We write the potential as
\begin{equation}
V(\phi) = \sum_{i=0}^K v_i (\phi/m_{{\rm Pl}})^i \,.
\end{equation}
The initial condition ranges that Kinney uses are given in terms of slow-roll 
parameters defined from the Hubble parameter; to obtain equivalent ranges on the 
coefficients of the potential we use the first-order slow-roll relations from 
Ref.~\cite{LPB}. This gives 
\begin{eqnarray}
v_1 &\simeq& -\sqrt{16\pi\epsilon_0} \,, \nonumber\\
v_2 &\simeq& 4\pi\left(\frac{\sigma_0}{2}+3\epsilon_0\right)\,,\nonumber\\
v_3 &\simeq& -\frac{8\pi^2}{3\sqrt{\pi\epsilon_0}}\left[6\epsilon_0^2+
(^2\lambda_{{\rm H},0})+\frac{3}{2}\epsilon_0\sigma_0\right]\,,\\
v_4 &\simeq& \pi^2\left[\sigma_0^2+8\epsilon_0\sigma_0+16\epsilon_0^2+
\frac{16}{3}(^2\lambda_{{\rm H},0}) 
+\frac{4}{3}\frac{(^3\lambda_{{\rm H},0})}{\epsilon_0}\right]\,\nonumber\\
v_5 &\simeq&
-\frac{8\pi^{5/2}}{3\epsilon_0^{1/2}}\left[\sigma_0(^2\lambda_{{\rm
H},0})+4\epsilon_0(^2\lambda_{{\rm H},0})+(^3\lambda_{{\rm H},0})
\right. \nonumber \\ && \left.
+(^4\lambda_{{\rm H},0})/(5\epsilon_0)\right]\,\nonumber\\
v_6 &\simeq& \frac{-2\pi^3}{45\epsilon_0^2}
\left[174\epsilon_0^2\sigma_0(^2\lambda_{{\rm H},0})
+168\epsilon_0^3(^2\lambda_{{\rm H},0})
\right. \nonumber \\ && \left.
+33\epsilon_0\sigma_0^2(^2\lambda_{{\rm H},0})
+40\epsilon_0(^2\lambda_{{\rm H},0})^2
+90\epsilon_0\sigma_0^2(^3\lambda_{{\rm H},0})
\right. \nonumber \\ && \left.
+360\epsilon_0^2(^3\lambda_{{\rm H},0})
+72\epsilon_0(^4\lambda_{{\rm H},0})
+12\sigma_0(^4\lambda_{{\rm H},0})
\right. \nonumber \\ && \left.
+4(^5\lambda_{{\rm H},0})\right]\,\nonumber  
\end{eqnarray}
Those relations are also used to evaluate the observables once the position of 
50 $e$-foldings is found.

\subsection{Results}

In Figure~\ref{fig3} we plot the results for each expansion for 6000
accepted initial conditions, replotting again the flow equations case
with that number of points for comparison. The same range has been
chosen for the observables in each case.

Each of the expansions is plotted to sixth-order. In order to check the 
convergence of the method, we have also analyzed the flow equations at 
fifth-order (to compare with Kinney \cite{K}) and eighth-order, and the other 
three 
methods at fourth-order. With the exception of the Pad\'e approximant for 
$\epsilon(\phi)$, discussed further below, there were no significant
changes in the distributions obtained indicating that reasonable 
convergence had occurred.

We see that there are significant differences between the 
models, though each does show some level of concentration in the
observable plane. Of the three new methods, the Taylor expansion of
the potential gives results closest to the flow equations, showing
indications of the same swathe of points with non-zero $r$, but not
however reaching to such high values. The classification of points is
very similar to the flow equations, with 90\% trivial points, and
almost all the remainder requiring backwards integration to achieve 50
$e$-foldings.

By contrast, the two $\epsilon(\phi)$ expansions give results which
are visually quite different. The Taylor series gives a diffuse
ensemble of points, with a preference for $n>1$ but covering a fairly
large fraction of the observable plane. The classification of points
is also quite different in this case, with a higher fraction of
points, about 15\%, giving 50 
$e$-foldings of inflation from the forward integration as compared to those 
requiring backwards integration (recall this method does not generate trivial 
points).

The Pad\'e approximant expansion of $\epsilon(\phi)$ gives a different
outcome again, with the separate classifications of points leading to
different groupings in the plane. The vast majority of the 
points shown correspond to backwards integration. The main feature at $n<1$ 
corresponds to $\epsilon(\phi)$ functions which
approach zero in the backwards integration, but do generate 50
$e$-foldings before reaching that point; these generate a similar structure as 
the flow 
equations. The grouping of points to the right of that at high $r$ corresponds 
to backwards integration
points, but this time to functions $\epsilon(\phi)$ which approach
one and give 50 $e$-foldings before that point. 
There is a third distinct grouping, mainly at $n>1$ and small $r$, corresponding 
to
points achieving 50 $e$-foldings in the forward integration, but it
contains very few points (2\% of the total).

However it is less easy to draw firm conclusions from the Pad\'e approximant, 
because we found the method is much less well converged than the others. When we 
ran this method at fourth-order rather than sixth-order, the same general 
structures were picked out, but the balance of points was quite different with 
most of the points lying in the right-hand set rather than the familiar flow 
equations swathe. By contrast, for the other methods the results were 
essentially unchanged between fourth-order and sixth-order.

\section{Conclusions}

We have investigated several ways of randomly generating sets of
inflation models, in order to compare their predictions in the
observational plane with those of the flow equations approach. We have
seen that the different methods, all of which are comparably well
motivated, give significantly different predictions.

In two of our new methods, we see hints of the structure seen in the
flow equations, but much less well defined. Models lying in that
region do seem particularly well suited to generating a sufficient
number of $e$-foldings, but the narrowness of the band appears to some
extent to be an artifact of the flow equations implementation. In
particular, a Taylor expansion of $\epsilon(\phi)$, which seems as
well motivated as the flow equations approach, does not reproduce such
a coherent structure in the observable plane.

%%%%%%%%%%%%%%%%%%%%%%%%%%%%%%%%%%%%%%%%%%%%%%%%%%%%%%%%%%%%%%%%%%%%%%%%
\begin{acknowledgments}
E.R. was supported by Conacyt and A.R.L. by PPARC. We thank Will
Kinney for useful discussions. E.R. is indebted to Martin Kunz
for his valuable help in Fortran and useful
discussions and comments. She also thanks Fernando Santoro, 
Micha\"{e}l Malquarti, Liam O'Connell, James Fisher, 
Diana Hanbury, Peter Thomas, David Rowley, Neil Bevis and Jon
Urrestilla for their kind help. 
\end{acknowledgments}
%%%%%%%%%%%%%%%%%%%%%%%%%%%%%%%%%%%%%%%%%%%%%%%%%%%%%%%%%%%%%%%%%%%%%%%%

%%%%%%%%%%%%%%%%%%%%%%%%%%%%%%%%%%%%%%%%%%%%%%%%%%%%%%%%%%%%%%%%%%%%%%%
\end{document}